\documentstyle[sprocl,psfig]{article}

\def\Journal#1#2#3#4{{#1} {\bf #2}, #3 (#4)}

\def\NPB{{\em Nucl. Phys.} B}
\def\PLB{{\em Phys. Lett.}  B}
\def\PRL{\em Phys. Rev. Lett.}
\def\PRD{{\em Phys. Rev.} D}
\def\PRC{{\em Phys. Rev.} C}
\def\ZPC{{\em Z. Phys.} C}
\def\PA{{\em Physica} A}
\def\AP{{\em Ann.Phys. (N.Y)}}

\def\be{\begin{equation}}
\def\ee{\end{equation}}
\def\bea{\begin{eqnarray}}
\def\eea{\end{eqnarray}}

\newcommand{\Od}{{\cal O}}

\newcommand{\tr}{\mbox{tr}}

\newcommand{\fpi}{f_\pi}

\newcommand{\ft}{f(t)}

\newcommand{\intcc}{\int_C \! d^4x}
\newcommand{\vxt}{(\vec{x},t)}

\begin{document}


\title{NONEQUILIBRIUM CHIRAL DYNAMICS AND EFFECTIVE LAGRANGIANS}

\author{A.G\'OMEZ NICOLA}

\address{Departamento de
F\'{\i}sica Te\'orica II,  U.Complutense, 28040
Madrid. SPAIN.\\E-mail: gomez@eucmax.sim.ucm.es}


\maketitle\abstracts{We review our recent work on
 Chiral Lagrangians out of thermal equilibrium, which are
  introduced to analyse the   pion gas
  formed after a Relativistic Heavy Ion Collision. Chiral Perturbation
  Theory is extended by letting $\fpi$ be time dependent and allows to
   describe explosive production of pions in parametric resonance.
This mechanism  could be relevant if hadronization occurs at the chiral phase
  transition.}

\section{Introduction}

In the last few years there has been a growing interest in
nonequilibrium  Quantum Field Theory, motivated by
  the experiments on Relativistic Heavy
Ion Collisions seeking the Quark-Gluon Plasma (QGP) and its properties.
 After the promising results obtained by SPS at CERN  the
 torch has been passed on to RHIC at the BNL. Roughly speaking, these
 experiments observe the final particle spectra, trying to extract
  information about the possible QGP formation and its
 subsequent expansion until freeze-out, during which the chiral phase
 transition and hadronization take place. Apart from  equilibrium
 properties 
 such as in-medium  masses and
decay widths or the phase diagram, nonequilibrium aspects of the
 expansion may also be relevant. An important example is the
 hadronization process, which is not yet fully understood. One of the
 scenarios suggested to explain the production of hadrons out of the
 expanding plasma is that of  Disoriented Chiral Condensates,
  where pions develop strong fluctuations
 after the chiral phase transition \cite{an89,rawi93}. In this
 context, two  mechanisms can give rise to pion production.
 The first is spinodal decomposition, which takes
place in an early stage after
 the transition, when it is reasonable to assume an initial supercooled
 state. Thus, long wavelength modes grow exponentially yielding strong
 pion correlations \cite{rawi93,bodeho95,cooper9596}. The second
 is
 parametric resonance, 
 where the oscillations of the $\sigma$ mode
 in a later stage yield explosive pion
 production.  The  resulting reheating process
 yields a final temperature $T_f\simeq$ 135 MeV compatible with the
 freeze-out of hadrons \cite{mm95}. Both approaches are complementary and have
 been studied numerically in great detail in the
 large-$N$ limit of the $O(N)$ model \cite{boyaetal96}.
Here we will discuss the
extension of Chiral Perturbation Theory (ChPT) 
to include nonequilibrium effects and its
application to pion production in parametric resonance \cite{agg99,agn01}.

\section{Nonequilibrium effective chiral models}

An effective model to describe QCD well below  the chiral scale
$\Lambda_\chi\simeq$ 1.2  GeV must be based on the chiral
Spontaneous Symmetry Breaking (SSB) pattern
  $SU_L(N_f)\times SU_R(N_f) \longrightarrow SU_V (N_f)$ where $N_f$
  is the number of light flavours. One example is the $O(N)$ model, which
   reproduces the chiral SSB only for
  $N=4$ ($N_f=2$) and, due to the strong interaction,
   is nonperturbative in the coupling constant, so that
  nonperturbative schemes such as large $N$ must be implemented.
An alternative for  low-energy QCD is  
ChPT \cite{we79,gale} which provides a well-defined expansion in
$p/\Lambda_\chi$ where $p$ is any meson energy. Such expansion is
renormalizable order by order. For instance, the $\Od(p^2)$
lagrangian, the familiar Non-Linear Sigma Model (NLSM), generates
$\Od(p^4)$ loops  whose divergences are absorbed by adding to
the NLSM a $\Od(p^4)$ lagrangian including all possible terms
compatible with the symmetries, with  undetermined coefficients, the
so-called low-energy constants (LEC), which can be fixed
with experimental information. Regarding nonequilibrium, the ChPT
scheme is best suited for a stage where the system is into the
broken phase. In that case, it provides us with the  advantages
commented above. For instance, it can be used both for $N_f=2,3$.
Therefore, as far as pion production is concerned, it will be
useful in the parametric resonance regime, whereas the $O(N)$
model is
  more adequate to describe the spinodal phase. The first step is
  then to extend the NLSM out of equilibrium. The
  approach we will follow is to perturb the system from an
  equilibrium state and analyze its subsequent evolution.
   Since the form of the lagrangian is dictated
  by the chiral symmetry, assuming spatial homogeneity and
  isotropy, the only way to introduce the perturbation to leading order is by
  letting  the pion decay constant $\fpi$ be time
  dependent \cite{agg99}. Thus, the $\Od(p^2)$ NLSM  nonequilibrium action
  becomes
\begin{equation}
S_2[U]=\intcc \ \frac{f^2(t)}{4} \ \tr \ \partial_\mu U^\dagger
\vxt
\partial^\mu U \vxt
 \label{nlsmne}
\end{equation}
where $\int_C$ indicates that the time integration runs along a
 Schwinger-Keldysh contour in the complex plane, which is the 
standard technique
  to formulate the non-equilibrium path integral. We fix the
  initial time to $t=0$, where the system is in 
thermal equilibrium at temperature
 $T_i$.  Thus, $f(t\leq 0)=f$ where $f=\fpi(1+\Od(p^2))$ and
 $\fpi\simeq$ 93 MeV.
  We restrict to $N_f=2$ so that the
 $SU(2)$ matrix
$U=[(f^2(t)-\pi^2)^{1/2} I + i\vec{\tau}\cdot\vec{\pi}]/\ft$ where
$\pi^a\vxt$ are
 the three pion fields. Note also that we work in the chiral limit, 
i.e, the $u,d$
masses vanish exactly, so that we have not included  explicit symmetry
breaking mass terms. One can then proceed with ChPT
as usual, expanding the action (\ref{nlsmne}) in pion fields,
which generates all possible interaction vertices. The 
chiral power counting  remains consistent if $\dot
f/f\simeq \Od (p)$, $\ddot f/f\simeq \Od (p^2)$ and so on. In this
sense, ChPT is more appropriate to describe a situation not far
from equilibrium such as parametric resonance. As explained
before, from the moment we consider loop diagrams the ChPT scheme
requires the $\Od(p^4)$ lagrangian, whose nonequilibrium version
is more involved to construct than the $\Od(p^2)$ NLSM in
(\ref{nlsmne}). For that purpose, it is useful to view the action
(\ref{nlsmne}) as an equilibrium NLSM in a spatially flat
Robertson-Walker (RW) metric background whose scale factor is just
$\ft/f$. This is easily achieved by rescaling the  pion
fields to $\pi^a f/\ft$ and allows to write the lagrangian to any
order by rising and lowering indices with the RW metric and including all
new tensor structures preserving the symmetry. The $\Od(p^4)$
lagrangian built in this way is given in \cite{agg99,agn01}. In
addition to the equilibrium lagrangian, it involves two new terms
and hence two new LEC, which  have been determined experimentally
in \cite{dole91}. These two terms are crucial since they allow to
renormalize new time-dependent infinities that were not present in
the equilibrium case. For instance, the ChPT one-loop diagrams
include tadpoles
    proportional to the equal time correlator
   $\int \frac{d^{d-1} k}{(2\pi)^{d-1}} G_0 (t,t,k)$,
   where $d$ is the space-time dimension and
$G_0 (t,t',k)=\langle T_C \pi_k (t) \pi_k (t') \rangle$ is the
pion two-point function.
   This quantity has a time-dependent divergent part which vanishes in 
   equilibrium.
    Using   dimensional regularization, these divergences can be
    absorbed in the new  $\Od(p^4)$  parameters  so that
    observables become finite and scale independent \cite{agn01}.

\section{Parametric Resonance in ChPT}

In the nonequilibrium $O(4)$ model, 
 $\langle\sigma\rangle$ is time dependent and  in the last
 stage of the nonequilibrium evolution it 
 oscillates around the constant vacuum. Thus,
 $\sigma\vxt=\sigma_0 (t)+\delta\sigma\vxt$ where $\sigma_0(t)$ is
 a time-dependent classical background which oscillates around
 $\sigma_0=f$ and $\delta\sigma$ includes quantum corrections,
 which are subleading. The oscillations transfer energy
 to the pion fields, yielding exponential growth of the pion correlator
 via parametric resonance \cite{mm95}. The connection with  ChPT is
 established through the observation that the NLSM corresponds, to
 leading order, to the $O(4)$ model with the constraint
 $\sigma^2+\pi^2=\sigma_0^2(t)$ and $\ft\sim\sigma_0 (t)$.
 Therefore, we will take
  \be
f(t)=f\left[1-\frac{q}{2}\sin Mt \right] \qquad (t>0) \label{ft}
 \ee

 It is important to bear in mind that using (\ref{ft}) is valid
 for times $t<t_{BR}$ where we denote by $t_{BR}$ the
 back-reaction time such as the pion correlations become typically
 $\langle \pi^2 \rangle (t_{BR})\sim f^2$. When that time scale is
 reached, (\ref{ft}) is no longer a solution of the equations of
 motion in the $O(4)$ model and  loop
 corrections in ChPT become of the same order as the tree level.
 It must be pointed out that   for $t\sim t_{BR}$, dissipation
 makes particle production stop\cite{boyaetal96}. 
Therefore, 
 the ChPT approach is able to account for nearly all the pion
 production before $t_{BR}$. Another point with is worth stressing is that
 to one loop in ChPT, the nonequilibrium chiral power counting
 demands $qM^2=\Od(p^2)$. That is, we are studying small
 oscillations, which in inflationary 
Cosmology is called narrow resonance. 
 The mass parameter $M$ does not need to
 be small and in fact numerically $M\sim m_\sigma$, although there
 is no need to invoke the $\sigma$ particle in the ChPT context.
 Once the assumptions and limitations of our approach are clear,
let us sketch how the parametric resonance works in ChPT. To
$\Od(\pi^2)$, the action (\ref{nlsmne}) yields the "free"
lagrangian, which now contains a time-dependent mass term, so
that, taking into account (\ref{ft}), $G_0$ satisfies 
\begin{equation} \left\{\partial_t^2 + k^2
-(qM^2/2)\sin Mt\right\} G_0 (t,t',k)=-\delta_C (t-t')
\label{lopropeq}
\end{equation}

The above differential equation has the form of the Mathieu
equation, whose solutions develops instability bands in momentum
space, centered at $k_n\simeq n M/2$ and of width $\Delta
   k_n=\Od (q^n)$. Therefore, to the order we are considering,
   i.e, $\Od(q)$,
   only the first band is resonant.

\begin{figure}[t]
\psfig{figure=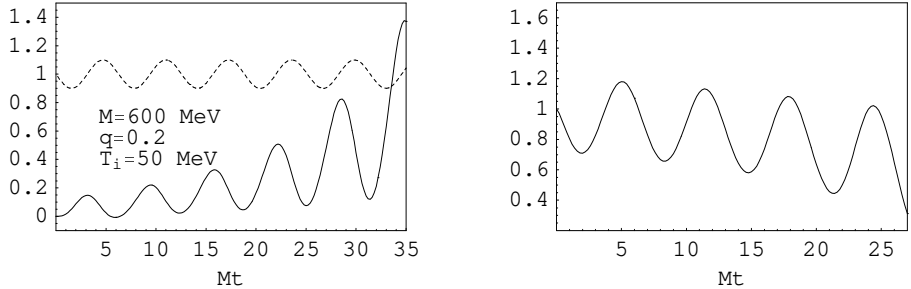,height=1.5in,width=4.7in} \caption{The
left curve is  $\Delta_{unst} (t)/\fpi^2$ where the dashed line is
the tree level contribution in (\ref{ft}). The right curve shows
 $\fpi^s (t)/\fpi(0)$ for the same parameters.
\label{fig:fpi}}
\end{figure}

The first observable one can calculate in one-loop ChPT  is $\fpi
(t) $, whose NLO corrections include  tree level
diagrams  coming from the $\Od(p^4)$ lagrangian and tadpoles
containing the exponentially growing  function 
\be \Delta_{unst} (t)=\frac{i}{2\pi^2}
\int_{\Delta k_1} \!\!dk k^2 \left[G_0 (t,t,k)-G_0 (0,0,k)\right]
\ee 

 We have
plotted $\Delta_{unst} (t)$ and $\fpi^s(t)$ in Figure
\ref{fig:fpi}, where $\fpi^{s,t}(t)$ correspond to the 
spatial and time components of the axial current \cite{agg99,pity96}. 
 The parameters have been chosen so that
$t_{BR}\simeq$ 10 fm/c, the typical plasma lifetime. We have
estimated $t_{BR}$ as
the time when $\Delta_{unst}$ equals the tree level contribution.
Observe that the $\fpi (t)$ oscillations are not damped, which  is
a reminder that we have not included the back-reaction properly
and should not worry us as far as pion production is concerned. We
also see that the central value decreases with time. This effect
can be interpreted as a reheating of the system (since $\fpi$
decreases with the temperature) which here leads
to a final temperature $T_f\simeq$ 125-140 MeV \cite{agn01},
compatible with hadronic freeze-out.

\begin{figure}[t]
\psfig{figure=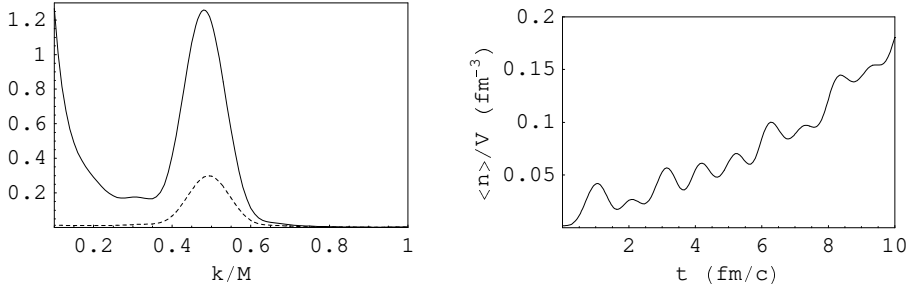,height=1.5in,width=4.7in} \caption{The
left solid
  curve is the time average of $n(k,t)$ from $t=0$ to $t_{BR}$ and
  the dashed line is that of $(k^2/M^2) n(k,t)$. The right curve is
  $\langle n(t)\rangle/V = (1/2\pi^2)
\int d k k^2  n(k,t)$. The parameters $M,q,T_i$ are the same as in
Figure \ref{fig:fpi}.
\label{fig:np}}
\end{figure}

The pion distribution function $n(k,t)$ and hence the pion number
may also be analysed in this context. First, one needs to provide
an appropriate definition for $n(k,t)$ at nonequilibrium since,
similarly to QFT in curved space-time, the vacuum state is
time-dependent. A consistent approach is to define it in terms of
the energy-momentum tensor expectation value,  using a suitable
point-splitting technique. In this way, one can  calculate the number of
massless pions to one loop in ChPT \cite{agn01}. After a careful
evaluation of all the NLO contributions it turns out
$n^{NLO}(k,t)=n^{LO}(k,t) + \Od(q^2)$. In Figure \ref{fig:np} we
have plotted the time average of $n(k,t)$ as well as the total
pion density as a function of time. The pion spectral function is
peaked around the resonant frequency
 $k\simeq$ 300 MeV, which could indeed be of importance for  
RHIC \cite{randrup00}. Note that the pion density
 grows without stop because the
back-reaction is neglected. However, as commented before, for
practical purposes we only need the number of  pions produced  by
the end of the expansion, for which we get  0.2 per fm$^3$. Our predictions
agree with the $\Od(4)$ model \cite{bodeho95,cooper9596,hiro}.

\section{Conclusions and Outlook}

Chiral Effective Models can be used to describe nonequilibrium
phenomena such as particle production, which are relevant for the
hadronization of a QGP. We have showed our results using ChPT,
which provides a systematic perturbative expansion in energies and
time derivatives. Explosive production of pions takes place when
$\fpi (t)$ oscillates in the parametric resonance regime. The
final pion spectra and pion density agree with recent
determinations within the $O(4)$ model. We believe that our
results are promising and provide useful techniques for future
theoretical analysis related to RHIC applications.

\section*{Acknowledgments}
 Financial
 support from CICYT, Spain,  projects AEN97-1693 and FPA2000-0956 is 
 acknowledged.

\end{document}